\documentclass{ws-procs9x6}

\begin{document}

\title{Parton intrinsic motion: unpolarized cross sections
and the Sivers effect in inclusive particle production\footnote{
\uppercase{T}alk delivered by 
\uppercase{F}.~\uppercase{M}urgia at the
``16th \uppercase{I}nternational \uppercase{S}pin \uppercase{P}hysics 
\uppercase{S}ymposium'', \uppercase{SPIN}2004, 
\uppercase{O}ctober 10-16, 2004, \uppercase{T}rieste, \uppercase{I}taly. }}
\author{U.~D'Alesio and F.~Murgia}
\address{Dipartimento di Fisica, Universit\`a di Cagliari and\\
INFN, Sezione di Cagliari, C.P. 170, 09042 Monserrato (CA), Italy}

\maketitle

\abstracts{
We present a detailed study,
performed in the framework of LO perturbative QCD with the inclusion 
of spin and $\bm{k}_\perp$ effects, of unpolarized cross sections
for the Drell-Yan process and for inclusive
$\pi$ and $\gamma$ production in
hadronic collisions, in different kinematical situations.
We find a satisfactory
agreement between theoretical predictions and experimental
data. This supports the study of spin effects and
transverse single spin
asymmetries (SSA) within the same scheme.
We then present results for SSA,
generated by the so-called Sivers effect,
in inclusive pion production in proton-proton
collisions.}

It is known that collinear pQCD, even at NLO,
often underestimates experimental results for inclusive
pion and photon production in hadronic collisions in the
central rapidity region and at moderately large $p_T$.
It has also been shown that intrinsic parton momentum
($\bm{k}_\perp$) effects reconcile in most cases theoretical
calculations with experimental results.
The role of $\bm{k}_\perp$ effects
has been also studied in the
context of  {\it polarized} high-energy inclusive particle production
at moderately large {\small $p_T$},
in particular at medium-large $x_F$.
Two noticeable examples are the SSA
observed in $p^\uparrow p\to\pi\,X$ processes, and the transverse
$\Lambda$ polarization measured in unpolarized hadronic collisions.
It was originally suggested by Sivers\cite{siv} and Collins\cite{coll}
that pQCD with proper inclusion of spin and
$\bm{k}_\perp$ effects in parton distribution/fragmentation
functions (PDF/FF) and in elementary dynamics
could be able to explain experimental results on {\small $A_N(\pi)$}.
This suggestion has been further extended in a number of 
subsequent papers.\cite{sico}
Recently, however, Bourrely and Soffer\cite{bour}
have claimed that most of the
experimental data on SSA cannot be explained within pQCD, on the basis
that collinear pQCD fails to reproduce the corresponding unpolarized
cross sections by 1-2 orders of magnitude.
As a matter of fact, in early pQCD-based approaches to SSA
a careful analysis of unpolarized cross
sections was not addressed.
Therefore, it was timely to perform a detailed and comprehensive study of 
unpolarized cross sections and SSA within the same pQCD approach and
at the same level of accuracy.
In this contribution we summarize the main results of this
program.
A detailed presentation and a complete list of references
can be found in Ref. \refcite{unp-kt}.

In order to study the inclusive production of moderately large
$p_T$ particles in high-energy hadron collisions, 
the process $AB\to C\,X$, we generalize, by including
spin and $\bm{k}_\perp$ effects, the
well-known collinear pQCD factorized expression for the
corresponding differential cross section:
\begin{eqnarray}
&&\!\!\!\!\!\!\!\frac{E_C\,d\sigma^{AB\to C\,X}}{d^3\bm{p}_C}
= \sum_{a,b,c,d}\int\!\!
dx_a\,dx_b\,dz\,\prod_i d^2\bm{k}_{\perp i}
\,\hat{f}_{a/A}(x_a,\bm{k}_{\perp a})\,
\hat{f}_{b/B}(x_b,\bm{k}_{\perp b}) \nonumber\\
\noalign{\vspace*{-4pt}}
&\!\!\!\times&\!\!\!\frac{\hat s}{x_ax_b s}
\frac{d\hat{\sigma}^{ab\to cd}}{d\hat{t}}(x_a,x_b,\hat{s},\hat{t},
\hat{u})\,\frac{\hat s}{\pi}\,\delta(\hat{s}+\hat{t}+\hat{u})\,
\frac{1}{z^2}\,J(z,|\bm{k}_{\perp C}|)
\,\hat{D}_{C/c}(z,\bm{k}_{\perp C})\,.\nonumber
\label{e:dsig}
\end{eqnarray}

The factors $\hat s/(x_a x_b s)$ and $J(z,|\bm{k}_{\perp C}|)$ are due
to $\bm{k}_\perp$ effects on the kinematics and become unity in
collinear pQCD with massless partons. The rest of the
notation should be obvious.\cite{unp-kt}
Additional contributions to the unpolarized
cross section are in principle possible;
however, it can be shown that they are negligible.\cite{unp-kt,kt-coll}
When considering SSA and Sivers effect only, to get the numerator
of the asymmetry (the denominator being twice the unpolarized cross
section)
one simply substitutes in this master formula the unpolarized
PDF with the corresponding Sivers function for the parton inside
the initial polarized hadron.
Notice that a formal proof of factorization with the inclusion of
$\bm{k}_\perp$ effects is still missing for $AB\to C\,X$ processes.
Moreover, a complete formal definition of spin and
$\bm{k}_\perp$ dependent PDF/FF, including their evolution
and universality properties is also missing; we also lack a 
consistent higher-twist
treatment including additional, unknown (higher-twist) PDF/FF
and possible quark-gluon correlations.
In fact, only ``enhanced'' higher-twist effects due to
$\bm{k}_\perp$'s are included in our approach:
i) the change of the
$x_{a,b}$, $z$ regions contributing mostly to the integrals in our
master formula can have a substantial effect, particularly
at large $x/z$ values, where PDF/FF vary rapidly;
ii) the partonic scattering angle in the
$AB$ c.m. frame might become much smaller than the hadronic
production angle, thus enhancing the moderately large $p_T$
production of particles.

Let us now summarize some basic details of
numerical calculations: {\it a}\,) We use a factorized, Gaussian-like
and flavour-independent behaviour for the $\bm{k}_\perp$-dependent
part of PDF/FF, e.g. $\hat f_{a/A}(x_a,\bm{k}_{\perp a}) =
f_{a/A}(x_a)\,(\beta^2/\pi)\>
\exp[-\beta^2\,k_{\perp a}^{\,2}]$ (where
$1/\beta=\langle\,k_{\perp a}^2\,\rangle^{1/2}$);
{\it b}\,) We calculate partonic cross sections at LO
in $\alpha_s$, with $\bm{k}_\perp$-modified
partonic invariants $\hat s$, $\hat t$, $\hat u$,
introducing a regulator mass,
$\mu=0.8$ GeV, so that
$\hat t \to \hat t -\mu^2, \hat u \to \hat u
-\mu^2, \hat s \to \hat s + 2\mu^2$;
{\it c}\,)
We use a unique scale for renormalization and factorization,
$Q=\hat{p}^*_T/2$, $\hat{p}^*_T$ being the transverse momentum
of the fragmenting parton in the partonic c.m frame;
{\it d}\,) We properly take into account NLO corrections ($K$-factors);
our LO numerical results
are always rescaled in the plots by a fixed (for a given curve) 
$K$-factor, estimated by using the numerical code INCNLL.\cite{aure}
{\it e}\,) We adopt MRST01 PDF, and Kretzer/KKP pion FF;
{\it f}\,) No complete best-fit procedure has been attempted at this stage;
therefore, in principle our results may be further improved.
\vspace*{-5pt}
\begin{figure}[ht]
\begin{center}
\hspace{-10pt}
\includegraphics[angle=-90,width = 7.3cm]{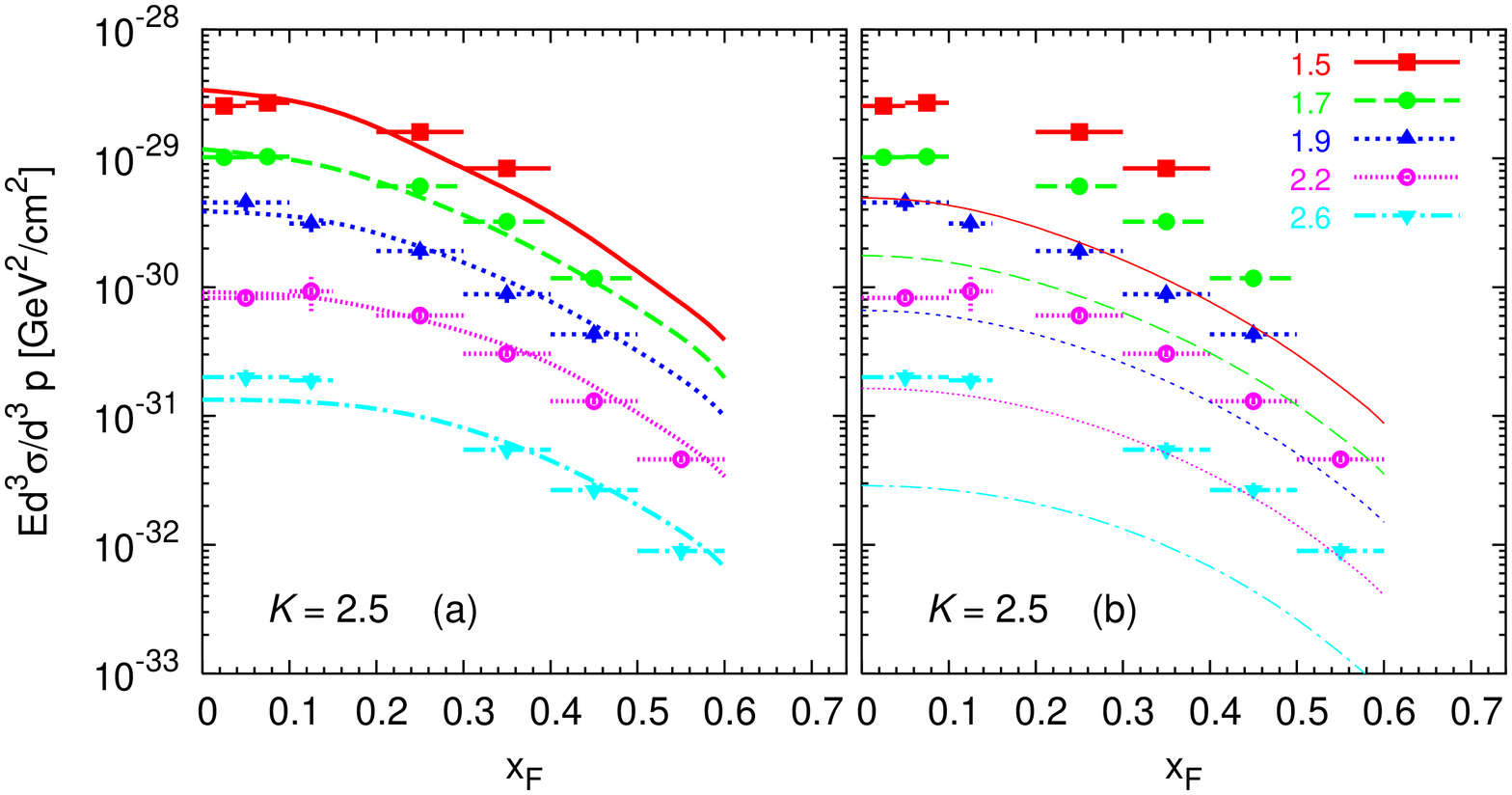}
\hspace*{-10pt}
\vspace*{20pt}
\includegraphics[angle=-90,width = 4.5cm]{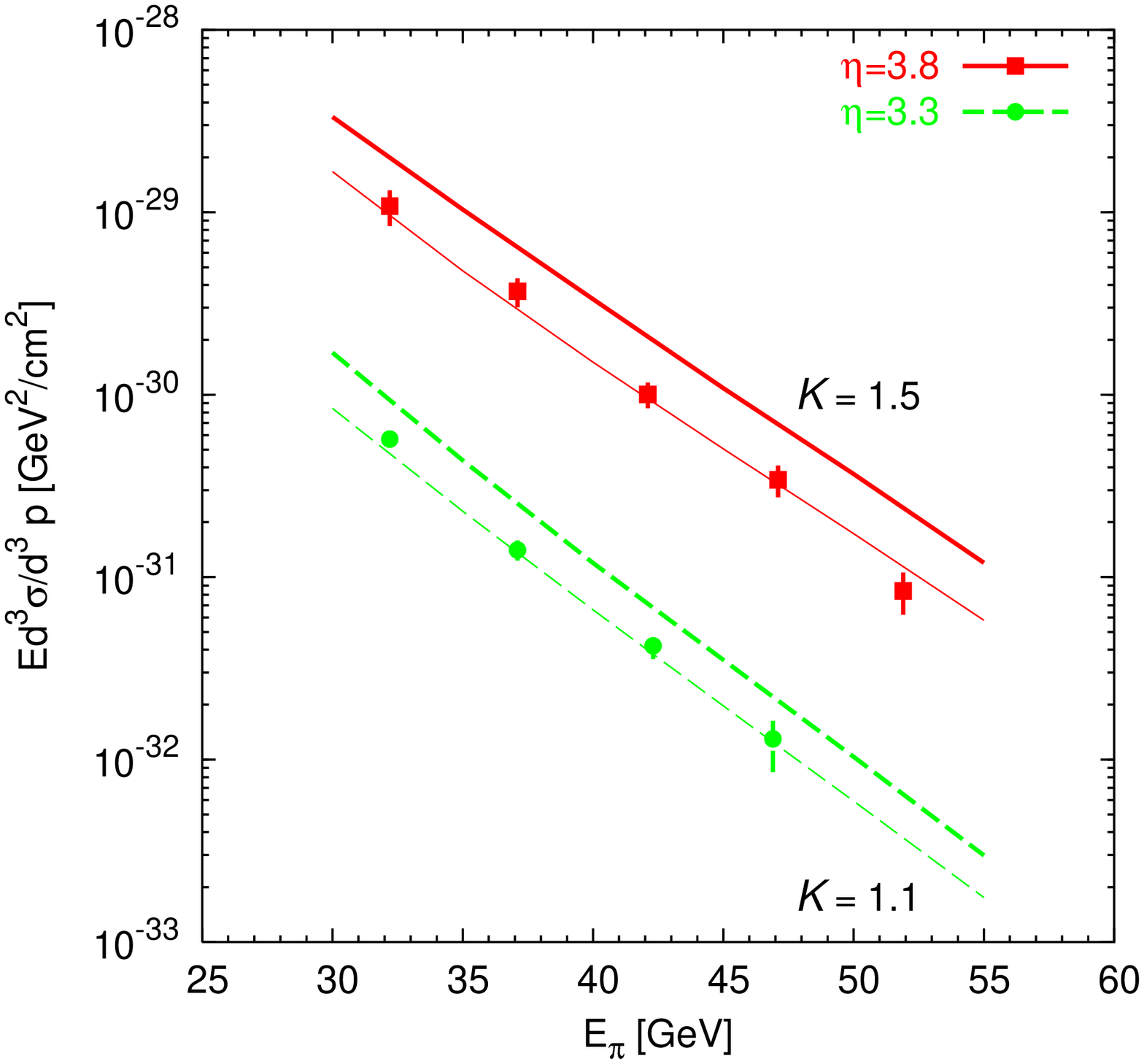}
\vspace*{-25pt}
\caption{Cross section for $pp\to\pi\,X$ process,
with (thick lines) and without (thin lines) $\bm{k}_\perp$
effects, compared
to FNAL\protect\cite{dona}
(left) and STAR\protect\cite{star03}
(right) data.}
\end{center}
\vspace*{-10pt}
\end{figure}

The following processes and
kinematical configurations have been considered:
1) $pp\to \mu^+\mu^-\,X$, in the region
$20 \le \sqrt{s} \le 60$ GeV, $5\le M \le 10$ GeV, 
$q_{T} < 3$ GeV/$c$ [used also to estimate
$1/\beta=\langle\,k_{\perp a,b}^2\,\rangle^{1/2}=0.8$ GeV$/c$ for PDF];
2) $pp\to \gamma\, X$, $\bar p p\to \gamma\, X$,
in the region
$20 \le \sqrt{s} \le 60$ GeV, $1.5\le p_T \le 10$ GeV/$c$,
$|x_{_F}|<0.4$, and
$\sqrt{s}\simeq 600$ GeV, $10\le p_T \le 80$ GeV/$c$;
3) $pp\to \pi\, X$ in the region
$20 \le \sqrt{s} \le 200$ GeV, $1.5\le p_T\le 14$ GeV/$c$,
$|x_{_F}|<0.8$ [used to estimate 
$1/\beta'=\langle\,k_{\perp\pi}^2\,\rangle^{1/2}$,
see Ref. \refcite{unp-kt}].
We can only present few representative results here:
in Fig. 1 we show the differential unpolarized cross
section for $pp\to\pi\,X$ in the kinematical configurations of
Ref. \refcite{dona} (left)
and of the STAR experiment at RHIC (right);
this gives a comparison between a calculation
similar to that of Ref. \refcite{bour} (thin lines) and our results
including $\bm{k}_\perp$ effects (thick lines).
In Fig. 2, we show $A_N(p^\uparrow p\to\pi\,X)$ in the kinematical
configurations of the E704 experiment at FNAL (left)
[used to fix the Sivers function] and
of the STAR experiment at BNL (right).
\begin{figure}[ht]
\begin{center}
\hspace{-10pt}
\includegraphics[angle=-90,width = 5.4cm]{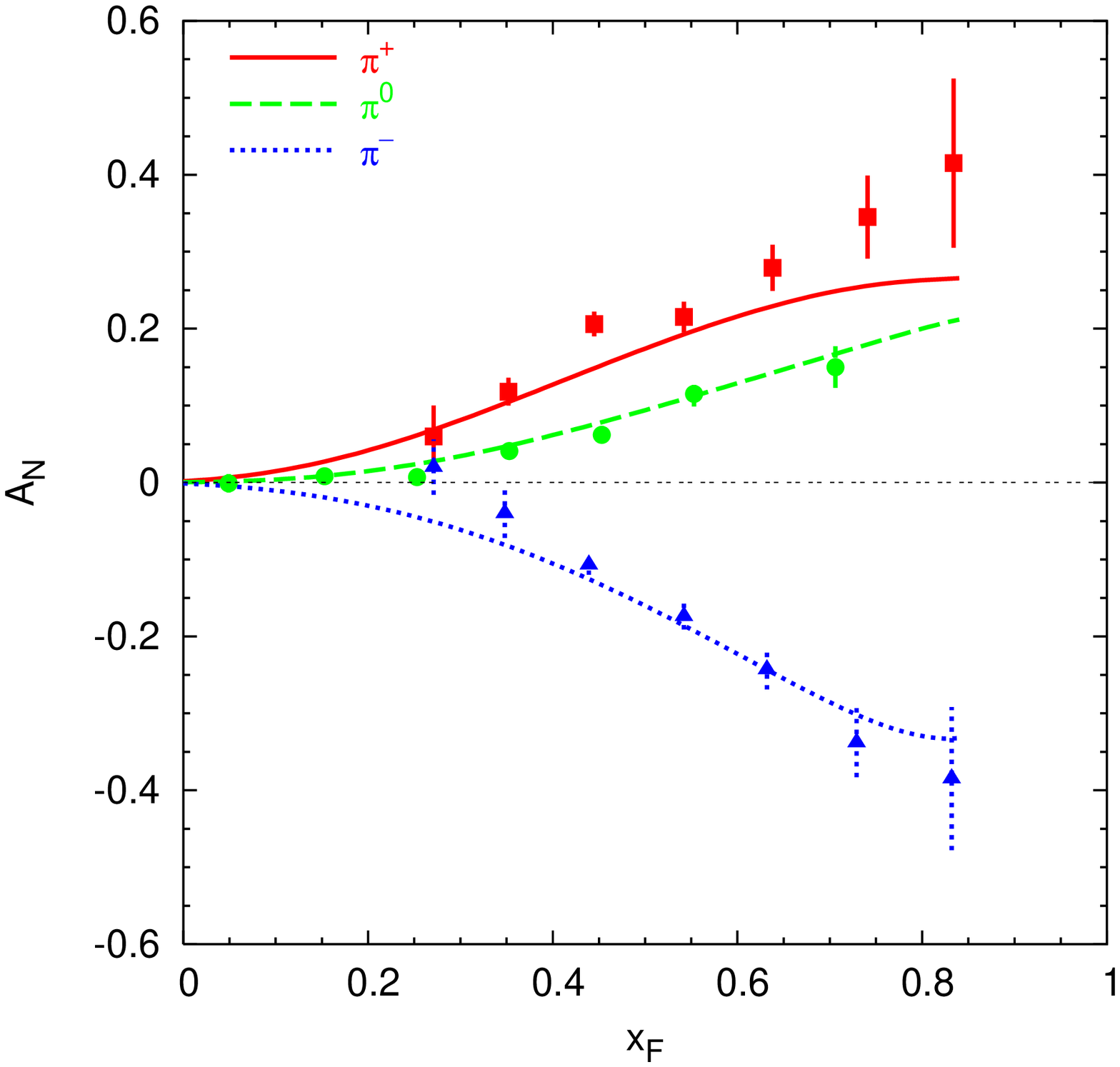}
\vspace*{20pt}
\includegraphics[angle=-90,width = 5.4cm]{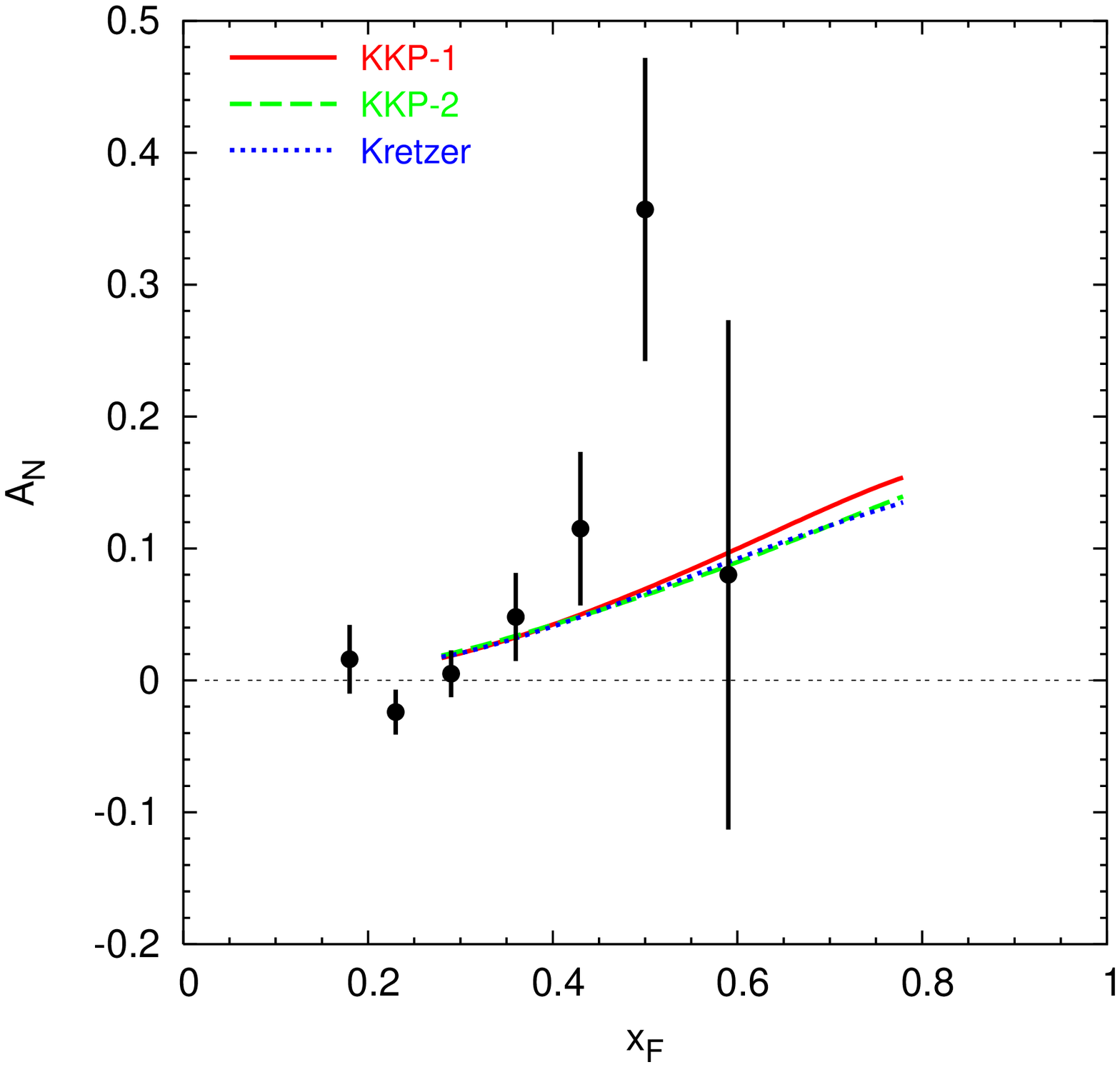}
\vspace*{-23pt}
\caption{$A_N(p^\uparrow p\to\pi\,X$) compared to
E704\protect\cite{e704}
(left) and STAR\protect\cite{star03}
(right) data.}
\end{center}
\vspace*{-10pt}
\end{figure}

In conclusion, our LO
approach (complemented with proper NLO $K$-factors) 
is in reasonable agreement with a large set of 
experimental data for unpolarized cross sections.
This gives support to the validity of the same approach in 
the study of SSA.
A detailed treatment of
$\bm{k}_\perp$ effects confirms all
main results and conclusions of our former studies on Sivers effect
performed keeping
only leading contributions (in $\bm{k}_\perp$) and using a simplified 
partonic configuration (see the first of Ref. \refcite{sico}).
Let us finally notice that our approach is in
principle able to reproduce most of the features of the 
experimental data available for $P_T^\Lambda(pp\to\Lambda^\uparrow\,X)$.
A combined analysis
of $P_T^\Lambda$ and unpolarized cross sections, like that performed here for
the Drell-Yan process and SSA in inclusive $\pi$ and $\gamma$ production,
is under way. 

We acknowledge partial 
support from ``Cofinanziamento MIUR-PRIN03''.


\begin{thebibliography}{00}
\bibitem{siv}     D. Sivers, {\it Phys. Rev.} {\bf D41}, 83 (1990);
                  {\it Phys. Rev.} {\bf D43}, 261 (1991)
\bibitem{coll}    J.C. Collins, {\it Nucl. Phys.} {\bf B396}, 16 (1993)
\bibitem{sico}    See, e.g., M. Anselmino, F. Murgia,
                  {\it Phys. Lett.} {\bf B442}, 470 (1998);
                  M. Anselmino, M. Boglione, F. Murgia,
                  {\it Phys. Rev.} {\bf D60}, 054027 (1999)
\bibitem{bour}    C. Bourrely, J. Soffer,
                  {\it Eur. Phys. J.} {\bf C36}, 371 (2004)
\bibitem{unp-kt}  U. D'Alesio and F. Murgia,
                  {\it Phys. Rev.} {\bf D70} 074009 (2004)
\bibitem{kt-coll} M. Anselmino, M. Boglione, U. D'Alesio, E. Leader, F. Murgia,
                  e-Print Archive: hep-ph/0408356
                  ({\it Phys. Rev. D}, in press)
\bibitem{aure}    P. Aurenche, {\it et al.},
                  {\it Nucl. Phys.} {\bf B297}, 661 (1988)
\bibitem{dona}    G. Donaldson, {\em et al.},
                  {\it Phys. Lett.} {\bf B73}, 375 (1978)
\bibitem{star03}  STAR Collaboration, J. Adams, {\em et al.},
                  {\it Phys. Rev. Lett.} {\bf 92}, 171801 (2004);
                  S. Heppelmann, contribution to the Transversity
                  Workshop, October 6-7, 2003, IASA, Athens, Greece
\bibitem{e704}    E704 Collaboration, D.L. Adams {\em et al.},
                  {\it Phys. Lett.} {\bf B345}, 569 (1995);
                  {\bf 261}, 201 (1991); {\bf 264}, 462 (1991)
\end{thebibliography}
\end{document}